\documentclass[conference]{IEEEtran}
\usepackage[latin9]{inputenc}
\usepackage{amsmath}
\usepackage{subfig}
\usepackage{graphicx}
\usepackage{balance}
\usepackage{xcolor}
\usepackage{multicol}
\usepackage{multirow}
\usepackage{algorithmic}
\usepackage{amssymb}
\usepackage{algorithm}

\ifCLASSINFOpdf
\else
\fi

\usepackage{stfloats}
\begin{document}

%
\title{In-Depth DCT Coefficient Distribution Analysis \\for First Quantization Estimation}


\author{\IEEEauthorblockN{Sebastiano Battiato}
\IEEEauthorblockA{University of Catania, Italy\\
battiato@dmi.unict.it}
\and
\IEEEauthorblockN{Oliver Giudice}
\IEEEauthorblockA{University of Catania, Italy\\
giudice@dmi.unict.it}
\and
\IEEEauthorblockN{Francesco Guarnera}
\IEEEauthorblockA{University of Catania, Italy\\
francesco.guarnera@unict.it}
\and
\IEEEauthorblockN{Giovanni Puglisi}
\IEEEauthorblockA{University of Cagliari, Italy\\
puglisi@unica.it}}


%


\maketitle

\begin{abstract}
\label{sec:abstract}
The exploitation of traces in JPEG double compressed images is of utter importance for investigations. Properly exploiting such insights, First Quantization Estimation (FQE) could be performed in order to obtain source camera model identification (CMI) and therefore reconstruct the history of a digital image. In this paper, a method able to estimate the first quantization factors for JPEG double compressed images is presented, employing a mixed statistical and Machine Learning approach. The presented solution is demonstrated to work without any a-priori assumptions about the quantization matrices. Experimental results and comparisons with the state-of-the-art show the goodness of the proposed technique.
\end{abstract}


%
\IEEEpeerreviewmaketitle

\section{Introduction}
\label{sec:intro}




Today, a typical life-cycle of a digital image is: (i) acquisition by means of a digital camera (often a smartphone); (ii) upload of the image to Social Network platforms or (iii) sending it through an Instant Messaging platform like Whatsapp or Telegram. The image has already gone through a double JPEG compression \cite{Giudice2017}. Thus, many information of the original image could be lost. If this image became the subject of a forensics investigation it is necessary to reconstruct its history \cite{farid2008digital} and then identifying camera model and camera source. At first, double compression must be detected (\cite{giudice20191,kee2011}), subsequently camera model could be inferred employing First Quantization Estimation (FQE) and finally camera source identification could be performed by means of comparison of Photo Response Non Uniformity (PRNU) (\cite{piva2013,stemm2013}).

In this paper a mixed statistical and Machine Learning approach for quantization factors estimation is presented. It employs a proper dataset of distributions for comparison, built without strong a-priori assumptions about the involved quantization matrices. Moreover, to reduce the computational complexity and the overall memory requirements, the proposed solution exploits Discrete Cosine Transform (DCT) distribution properties (i.e., AC coefficients are usually characterized by Laplacian distribution). Experimental results and comparisons demonstrated the robustness of the method.

The remainder of this paper is organized as follows: in Section~\ref{sec:related} related works are discussed; Section~\ref{sec:notation} introduces the reader to the JPEG notation; Section~\ref{sec:method} describes the proposed novel approach with discussion on its parameters in Section~\ref{sec:paramenter}. Experimental results are reported in Section~\ref{sec:results} and finally Section~\ref{sec:conclusion} concludes the paper.

\section{Related Works}
\label{sec:related}

First Quantization Estimation is extremely useful for forensics investigations giving hints about the history of a digital image. Many solutions were proposed in recent years. At first research activity was involved only in estimating the first quantization factors when the information was lost due to format change (JPEG to Bitmap). Fan and De Queiroz  in \cite{fan2000maximum}, \cite{fan2003identification} described a method to determine if an image in Bitmap format has been previously JPEG-compressed and thus they estimated the quantization matrix used. They firstly attacked this problem in an easier scenario of single compression and format change useful to understand JPEG artifacts and attract the attention of researchers  until today \cite{yang2020clustering}.


A first robust technique for FQE was proposed by Bianchi et al. (\cite{bianchi2012image,bianchi2011improved,bianchi2011detection}). They proposed a method based on the Expectation Maximization algorithm to predict the most probable quantization factors of the primary compression over a set of candidates. Other techniques based on statistical consideration of DCT histograms were proposed by Galvan et al. \cite{galvan2014first}. Their technique work at specific conditions on double compressed images exploiting the a-priori knowledge of  monotonicity of the DCT coefficients. Strategies related to histogram analysis and filtering similar to Galvan et al. were introduced until these days (\cite{dalmia2016first,yao2020jpeg}). Lately, insights used for steganography detection were exploited by Thai et al. \cite{thai2019estimation}: even if they achieved good results in terms of overall accuracy, they work only at specific combinations between first and second compression factors, avoiding the estimation when quantization factors are multiples.

Given the big amount of data available today, it is easy to figure out that the problem could be evaluated by means of modern Machine Learning approaches. J. Luk\'{a}\v{s} and J. Fridrich in \cite{lukavs2003estimation} introduced a first attempt exploiting neural networks, furtherly improved in \cite{varghese3detection} with considerations on errors similar to \cite{galvan2014first}. Convolutional Neural Networks (CNN) were recently also introduced in many papers \cite{BARNI2017153,uricchio2017localization,Wang2016}. CNNs have demonstrated to be incredibly powerful in finding invisible correlations on data, but they have been also very prone to what in Machine Learning is called overfitting: the situation in which the network modelled something that is not general enough to represent reality. Machine Learning techniques, and in this special case CNNs are strictly related to the dataset on which they trained. Accuracy and usability of these techniques have to be proved in "wild" conditions. Even though, state-of-the-art results are obtained recently by Niu et al. \cite{8945385} where top-rated results are reported for both the aligned and not-aligned FQE scenarios.

The proposed approach, by means of a minimal set of parameters and a reference dataset, aims at surpassing the limits of both CNN-based solutions and state-of-the-art analytical ones. This could be done employing statistics on DCT coefficients in such a way to remove the need of a training phase and without falling in the limits of analytical methods, full of a-priori assumptions.

\begin{figure}[t]
\centering
     
     \centering
     \captionsetup[subfloat]{labelformat=simple , labelsep=period}
     \subfloat[$M_i$\label{subfig:costant_table}]{%
       \framebox{\includegraphics[width=0.35\linewidth]{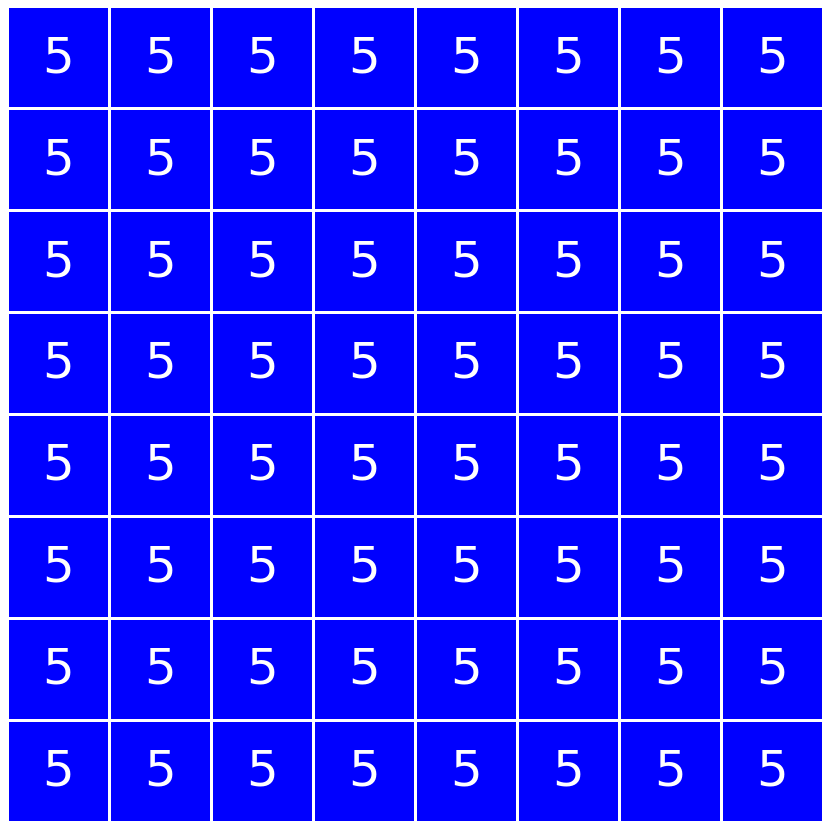}}
     }
     \hfill
     \centering
     \subfloat[Standard matrix $QF=90$\label{subfig:standard_table}]{%
       \framebox{\includegraphics[width=0.35\linewidth]{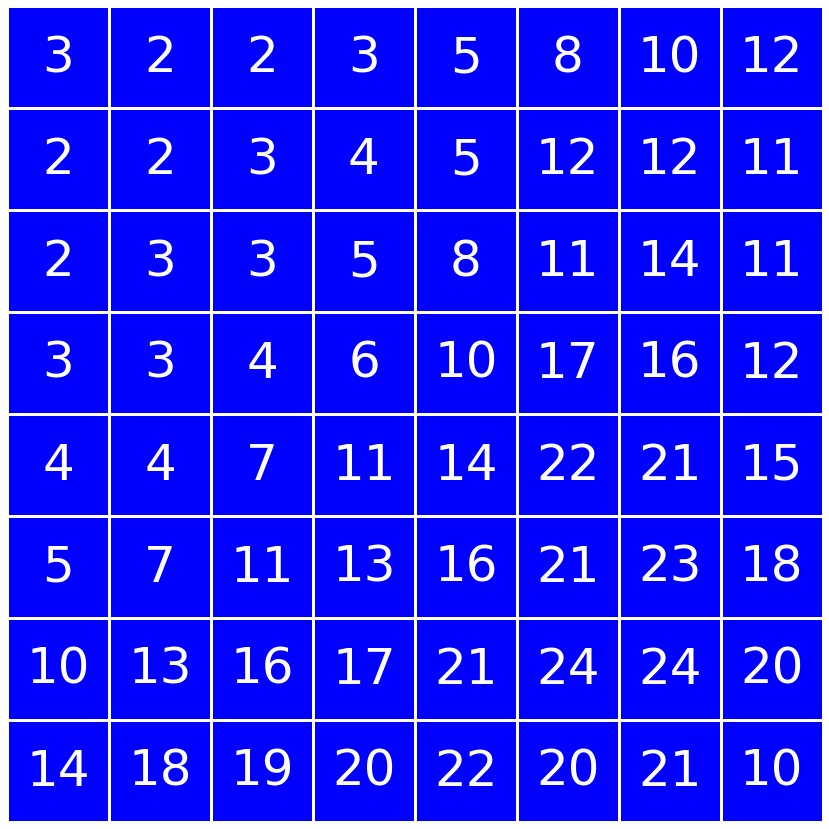}}
     }

     \caption{Example of constant matrix $M_i$ with $i=5$ (a), and the standard matrix corresponding to quality factor $QF=90$ (b). }
     \label{fig:quantization_matrices_example}
   \end{figure}

\section{JPEG Notation}
\label{sec:notation}

Given a raw image $I$, JPEG compression~\cite{wallace1991jpeg} could be defined as $I'=f_Q(I)$, where $I'$ is the JPEG compressed image and $Q$ is the $8 \times 8$ quantization matrix composed by the quantization factors $q_i \in \mathbb{N} $ with $i \in \{1,2,......,64\}$. Firstly $f_Q(I)$ converts $I$ from the RGB color space to the YCbCr ones, then divides it in $8 \times 8$ non-overlapping blocks and applies the DCT. Finally, each $8 \times 8$ block is divided by $Q$ pixel by pixel, rounded (losing information) and then encoded. In the analysis presented in this paper, only Y channel (luminance) will be considered.
Let's define $I''=f_{Q_2}(f_{Q_1}(I))$ a JPEG double compressed image, where $Q_1$ and $Q_2$ are the quantization matrices employed for the first and for the second compression respectively. 

The JPEG quality factor is defined as an integer $QF \in \{1,2,......,100\}$ referred to the quantization matrix $Q$ and describes the loss of information where $QF=100$ and $QF=1$ represent the minimum and maximum level of loss respectively. Various JPEG libraries could slightly change the implementation details. In this paper, we will refer to $QF$ as the standard quantization matrix defined by JPEG for a specific quality factor. $QF_i$ is defined as $QF$ used for the $i$-th JPEG compression.

In this paper, we denote $h_i$ the distributions of the $i$-th DCT coefficients in all the $8 \times 8$ blocks of $I''$. Furthermore, we define as $q1_1,q1_2,....,q1_{k}$ the $k$ quantization factors in zig-zag order of $Q_1$ which are the objective of the FQE, referring to $q1$ and $q2$ as the general quantization factor used in the first and in the second compression respectively. Finally, for testin purposes, we refer to $q1_{max}$ as the maximum value between all the quantization factors to be predicted (in Fig. \ref{subfig:standard_table} $q1_{max}=5$ considering $q1_1,q1_2,.....q1_{k}$ in zig-zag order with $k=15$).

\section{FQE through Comparison}
\label{sec:method}

\subsection{Retrieval distributions}
\label{sec:dataset}

The main aim of this work is the design of a solution able to properly exploit information contained in double JPEG compressed images without being affected by the typical limits of Machine Learning solutions. Specifically, in the state-of-the-art, many FQE methods have some restrictions in terms of quantization tables; for example Niu et al. \cite{8945385} provide  Neural Network models working only for standard matrices as $Q_2$ (with $QF_2=80/90$). Moreover, another contribution of the proposed method is the capacity to work in challenging conditions (i.e., custom matrices). 

The first (and most important) part is the generation of the reference data to be exploited in comparisons for the FQE. Starting from the RAISE dataset \cite{dang2015raise}, composed by $8157$ high resolution images in TIFF format (uncompressed), a patch of $64\times64$ pixels was extracted from the center. Hence, every crop was compressed two times, employing all the combinations of constant matrix $M_i$ (like the one shown in Figure \ref{subfig:costant_table}), with $i \in \{1,2,..,q1_{max}\}$ for the first and the second compression. Thus, generating $8157 \times 22 \times 22 = 3.947.988$ images. This dataset will be employed as a reference for comparisons, containing images double compressed by means of only those combinations of $M_i$ which do not belong necessarily to a specific example of real quantization matrices  (e.g., standard ones). This latter feature makes the overall approach generalizable on every kind of JPEG double compressed image in the aligned scenario. Moreover, it is worth noting that $q1_{max}=22$ is a good trade-off between the amount of the data to be generated ($22 \times 22$ combinations) and the maximum quantization factor to be estimated.

The comparison between the JPEG double compressed image under analysis and the generated reference dataset will be done employing the DCT coefficient distributions $h_i$. For this reason, the first $k$ $h_i$ were computed and then inserted into different sub-datasets identified by the couple $\{q1,q2\}$.

The insights described by Lam et al. \cite{lam2000mathematical} have shown the usefulness of Laplacian distribution (\ref{eq:laplacian}) over the years (\cite{farinella2015representing,hou2011image}), then we fitted the distributions $h_i$ on them, computing $\mu$
and $\beta$ to sort the sub-datasets:
\begin{equation}
	\label{eq:laplacian}
	\begin{split}
	f(x) =  {1\over2\beta}exp\Biggl({-|x-\mu| \over \beta} \Biggl)
	\end{split}
\end{equation}

The final dataset is then composed as follows:
\begin{itemize}
    \item  $DC_{dset}$: \textit{DC} distributions splitted for every possible couple $\{q1,q2\}$ sub-datasets, and sorted by $\mu$;
    \item  $AC_{dset}$: \textit{AC} distributions splitted for every possible couple $\{q1,q2\}$ sub-datasets, and sorted by $\beta$;
\end{itemize}


\subsection{Quantization Factor Estimation}
\label{section:quantization_estimation}

To estimate the first $k$ quantization factors of the first quantization matrix $Q_1$, namely \{$q1_1, q1_2, ......, q1_k$\},given $I''$,  the overall pipeline is summarized in Algorithm~\ref{algo:algo1}.


The estimation is done for every single $q1_i$ with $i\in\{1,2,.....,k\}$. Firstly, we extract the $h_i(I'')$ from $I''$ employing the LibJpeg C library~\footnote{\url{https://github.com/LuaDist/libjpeg}}, avoiding to further add truncation and rounding error. $h_i(I'')$ is then fitted on Laplacian distribution in order to extract $\mu$ and $\beta$ which are then used to seek the range of candidates from the reference dataset. The usage of $\mu$ and $\beta$ makes the number of candidates constant thus the computational cost to search the most similar distributions in the reference dataset is independent to its cardinality.

A JPEG file contains the quantization matrix used in the last compression; in our case $Q_2$ and then all the $q2_i$, allowing the selection of a specific  sub-dataset. For every sub-dataset selected in this way  \{$q_j$,$q2_i$\} with $j \in \{1,2,...,q1_{max}\}$, we extract a range of elements $D_{j,q2_i}(\mu,\beta)$ with the most similar values of $\mu$ for $DC_{dset}$ and $\beta$ for $AC_{dset}$ and then we compare $h_i(I'')$ with those elements using $\chi^2$ distance:

\begin{equation}\label{equation:chi_squared_distance}
{\chi}^2({x},{y})=\sum_{i=1}^{m} {(x_i - y_i)^2}/(x_i + y_i)
\end{equation}

where $x$ and $y$ represent the distributions to be compared.

For every sub-dataset $D_{j,q2_i}$, we select the lowest distance $d_{i,j}$ obtaining $q1_{max}$ distances. The minimum distance $d_{i,j}$, $j\in\{1,2,....,q1_{max}\}$ indicates the corresponding sub-dataset and then the predicted $q1$ for the current $i$.

\subsection{Regularization}
\label{subsec:method_regularization}

The $d_{i,j}$ distances obtained as described in previous Section, show that a strong minimum is not always present at varying of $j$. Sometimes, the information contained in $h_i(I'')$ does not clearly allow the discrimination among the possible $q1_i$ candidates. To overcome this, data coming from neighbors DCT coefficients can be exploited. Specifically, starting from the empirical hypothesis that a generic $q1_i$ value is usually close to $q1_{i-1}$ and $q1_{i+1}$, instead of estimating each coefficient independently, three consecutive elements in zig-zag order are considered. For example, if $k=15$, $13$ triplets ($q1_{i-1}$, $q1_{i}$, $q1_{i+1}$, $i=2,\dots,14$) can be identified. To estimate a single triplet $q1_{i-1}$, $q1_{i}$, $q1_{i+1}$, a score is associated to each possible $q1$ combination. Thus, considering $q1_{max}=22$ as the maximum $q1$ value, $22\times22\times22$ $q1$ combinations are taken into account. A proper score $S$ is then obtained by a weighted average between a data term ($C_{data}$) and a regularization term ($C_{reg}$) as follows:
\begin{equation}\label{equation:regularization}
S = w{C_{data}}+(1-w){C_{reg}}
\end{equation}
where $w \in [0,1]$, $C_{data}$ is the normalized sum of the three $d_{i,j}$, and $C_{reg}$ is the regularization term that tries to minimize the difference among the considered triplet. Further details about $C_{reg}$ setting will be provided in Section~\ref{par_reg}.

\begin{algorithm}[H]
    \centering
    \caption{The Proposed FQE Technique}\label{algo:algo1}
\begin{algorithmic}[1]
 \renewcommand{\algorithmicrequire}{\textbf{Input:}}
 \renewcommand{\algorithmicensure}{\textbf{Output:}}
 \REQUIRE double compressed image $I''$
 \ENSURE  \{$q1_1, q1_2, ......, q1_k$\}
 \\ \textit{Initialization} : $k$, $q1_{max}$
 \FOR {$i = 1$ to $k$}
  \IF {($i = 1$)}
  \STATE $D : DC_{dset}$
 \ELSE
  \STATE $D : AC_{dset}$
 \ENDIF
 \STATE $h_i$  : distribution of $i$-th DCT coefficient
 \STATE $\mu$, $\beta$  : $\mu$, $\beta$ fitted on Laplacian $h_i$ 
 \STATE $q2_i$ : quantization factor of $Q2$ for $i$-th DCT
 \FOR {$j = 1$ to $q1_{max}$}
 \STATE $D_{j,q2_i}$ : sub-dataset $(q1,q2)$ with $q1=j$, $q2 = q2_i$
 \STATE $D_{j,q2_i}(\mu,\beta)$ : sub-range with most similar $\mu,\beta$
 \STATE $d_{i,j}$ : lower $\chi^2$ distance between $h_i$ and $D_{j,q2_i}$
 \ENDFOR
 \STATE $q_i : argmin\{d_{i,j}\}$, $j\in\{1,2,....,q1_{max}\}$
 \ENDFOR
 \STATE regularize(\{$q1_1, q1_2, ......, q1_k$\})
 \RETURN \{$q1_1, q1_2, ......, q1_k$\} 
 \end{algorithmic} 
  \end{algorithm}

\section{Parameters setting}
\label{sec:paramenter}
The various parameters introduced in the Section~\ref{sec:method}, were set by means of a validation dataset $D_V$ composed by $8157$ images $64\times64$ pixels, cropped at random position from RAISE \cite{dang2015raise} original images. 
\begin{figure}[t]
\centering
     \hfill
    
      \includegraphics[width=\linewidth]{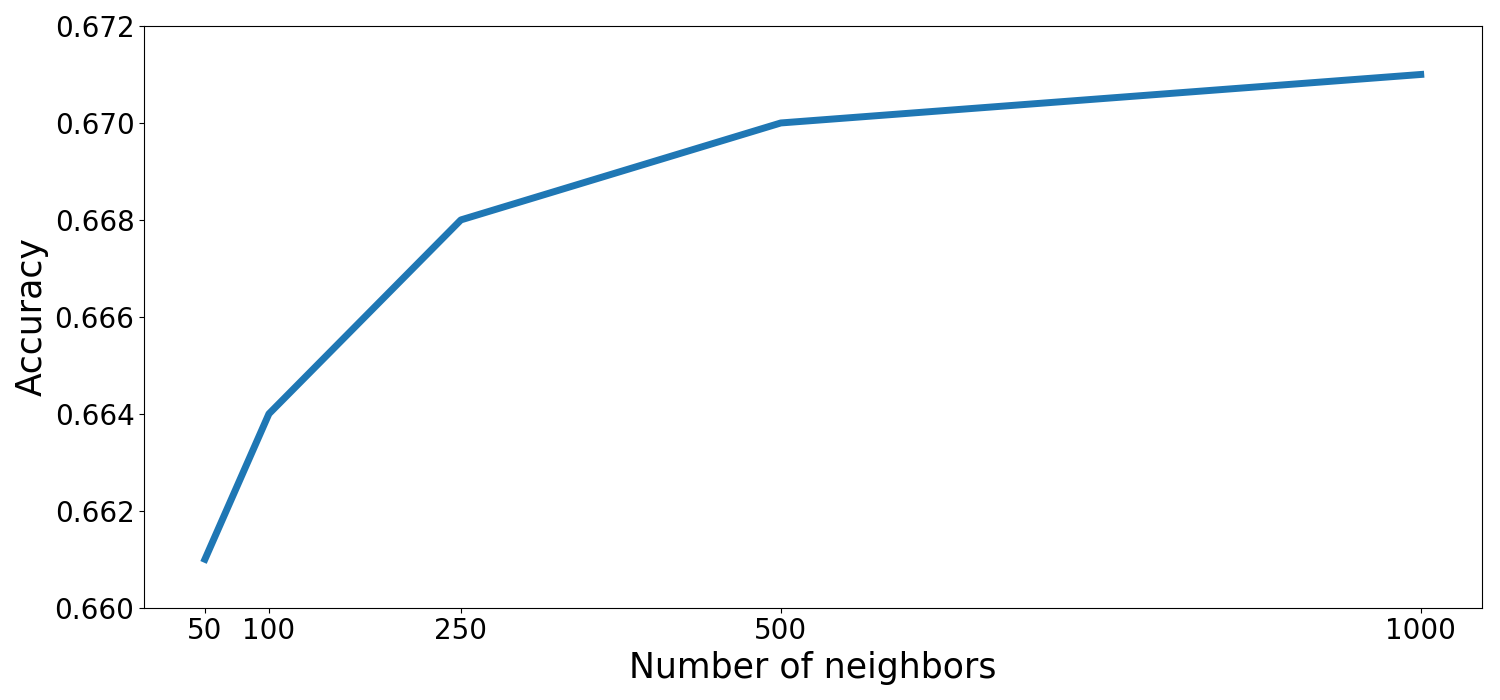}

     \caption{Accuracy of the proposed approach w.r.t. the number of neighbors considered for comparison.}
     \label{fig:accuracy_nearest}
   \end{figure}
\subsection{Clustering}
Fixed $q2$, the comparison dataset is composed by $q1_{max}$ different sub-datasets. In order to limit the overall computational complexity, a smart comparison strategy considering only a limited number of $h_i$ exploiting $\mu$ and $\beta$ values has been employed. We tested the method with $D_V$ using a sub-range of $50$, $100$, $250, 500$ and $1000$ elements for each sub-dataset (see Fig. \ref{fig:accuracy_nearest}). It is worth noting that the value considered in the proposed solution ($1000$), represents a viable trade-off between accuracy and computational cost with respect to the full search solution.

\subsection{Regularization}
\label{par_reg}

\begin{figure*}[t]
\centering
\captionsetup[subfloat]{labelformat=simple , labelsep=period}
     \subfloat[$C_{reg1}$\label{subfig:f1}]{%
       \includegraphics[width=0.32\linewidth]{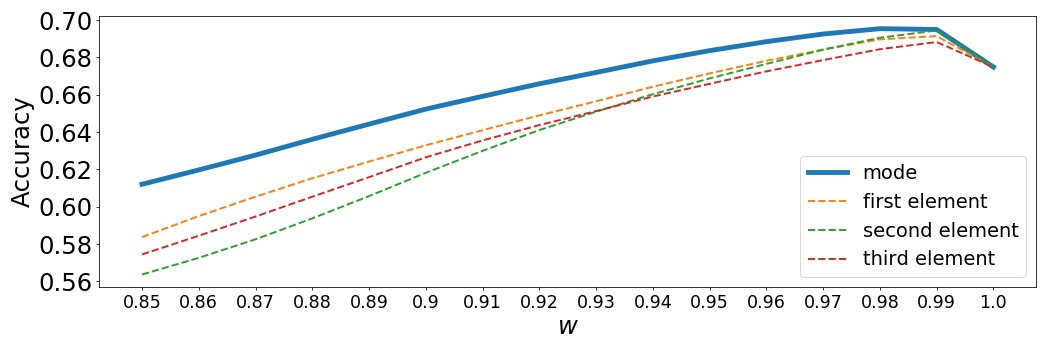}
     }
     \hfill
     \subfloat[$C_{reg2}$\label{subfig:f2}]{%
      \includegraphics[width=0.32\linewidth]{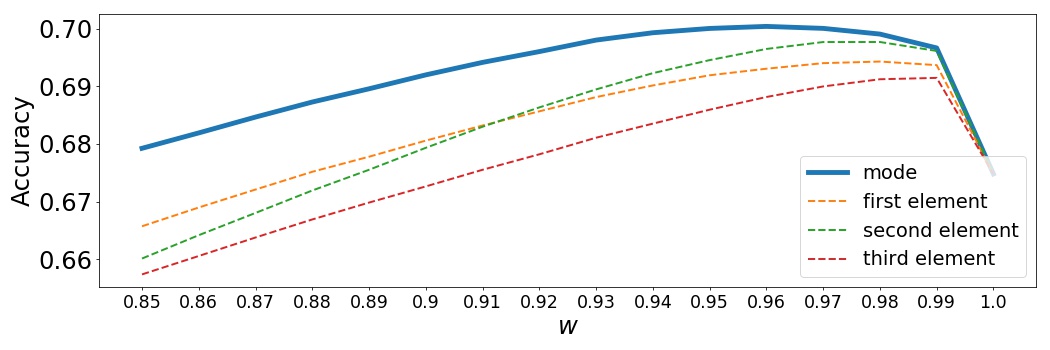}
     }
     \hfill
     \subfloat[$C_{reg3}$\label{subfig:f3}]{%
       \includegraphics[width=0.32\linewidth]{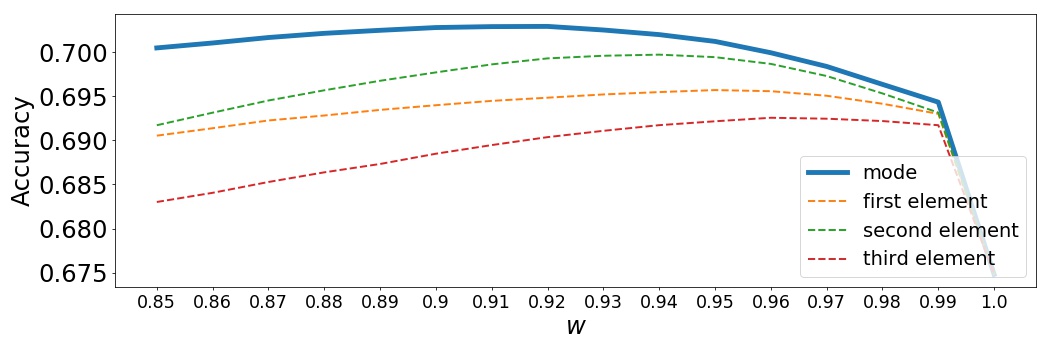}
     }

     \caption{a, b and c describe the accuracies obtained for different values of the regularization parameter $w$ employing the three equations (\ref{eq:f1}), (\ref{eq:f2}), (\ref{eq:f3}) respectively. Every plot shows the accuracies obtained considering the first, the second and the third element of the triplet of the analyzed quantization factors and the related arithmetical mode.}
     \label{fig:parameters_regularization}
   \end{figure*}

\begin{figure}[t]
\centering
     \hfill
     
      \includegraphics[width=\linewidth]{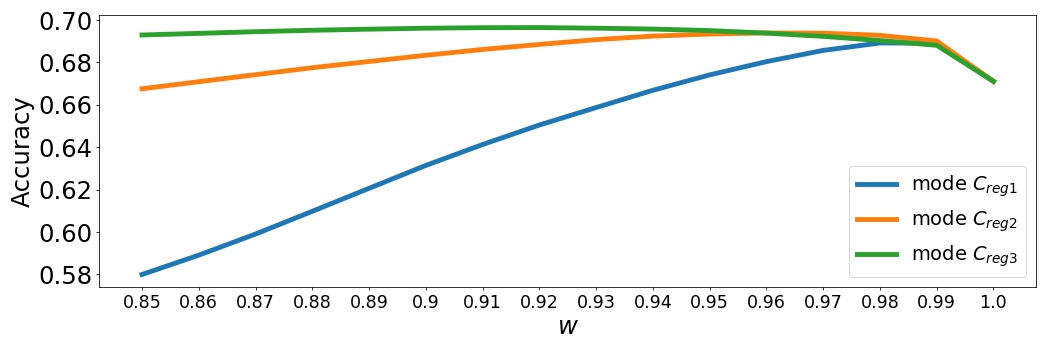}

     \caption{Comparison between the modes of Fig. \ref{subfig:f1}, \ref{subfig:f2}, \ref{subfig:f3}}
\label{subfig:mode_f1_f2_f3}   
\end{figure}

In order to cope with the lack of information contained in some $h_i(I'')$, a regularization step was carried out. Triplets of the nearest $q1_i$ were estimated simultaneously, and several $C_{reg}$ functions have been investigated: 

\begin{equation}\label{eq:f1} C_{reg1}=\frac{(c_i-c_{i-1})+(c_i-c_{i+1})}{2} \end{equation}\\
\begin{equation}\label{eq:f2} C_{reg2}=\frac{(c_i-c_{i-1})+(c_i-c_{i+1})}{2\sqrt(c_i)}\end{equation}\\
\begin{equation}\label{eq:f3} C_{reg3}=\frac{(c_i-c_{i-1})+(c_i-c_{i+1})}{2c_i}\end{equation}\\

where $c_{i-1}$, $c_{i}$, $c_{i+1}$ are $q1$ candidates related to three consecutive quantization factors in zig-zag order. This regularization step provides multiple estimations for each single $q1_i$. For example the estimations of $q1_3$ can be found on $3$ different triplets: ($q1_1$, $q1_2$, $q1_3$), ($q1_2$, $q1_3$, $q1_4$) and ($q1_3$, $q1_4$, $q1_5$). The $3$ strategies could then be tested. Figures \ref{subfig:f1}, \ref{subfig:f2} and \ref{subfig:f3} show the accuracies employing the corresponding $C_{reg}$, at varying of weights (see (\ref{equation:regularization})) and $q1_i$ selection strategies. As suggested in Fig. \ref{subfig:mode_f1_f2_f3} which shows the modes of eq. (\ref{eq:f1}), (\ref{eq:f2}), (\ref{eq:f3}), we chose the use of eq. (\ref{eq:f3}) with $w=0.92$.

\section{Experimental results}
\label{sec:results}

The difficulty of the first quantization factor estimation task is proportional to the relative position in the quantization matrix, as reported in literature. For this reason and for the sake of comparison with the state-of-the-art, we set $k=15$, although our method has not this specific limitation on $k$.

\subsection{Comparison tests}

 \begin{table}[t]
	\centering
	\resizebox{\linewidth}{!}{
	\renewcommand{\arraystretch}{1.3}
		\begin{tabular}{c||c||c||c||c||c||c||c||c||c||c||c||c}
			\hline
			 \textbf{$QF_1$} & \multicolumn{6}{c ||}{\textbf{$QF_2=90$}} & \multicolumn{6}{c}{\textbf{$QF_2=80$}} \\
			\hline \hline
			& \textit{Our} & \textit{Our Reg.} & \textit{\cite{bianchi2012image}} & \textit{\cite{galvan2014first}} & \textit{\cite{dalmia2016first}} & \textit{\cite{8945385}}   	& \textit{Our} & \textit{Our Reg.} & \textit{\cite{bianchi2012image}} & \textit{\cite{galvan2014first}} & \textit{\cite{dalmia2016first}} & \textit{\cite{8945385}} \\
			\hline \hline
			55 & 0.76 & \textbf{0.77} & 0.53 & 0.52 & 0.45 & 0.00 & 0.55 & \textbf{0.58} & 0.36 & 0.37 & 0.37 & 0.24\\
			\hline
			60 & \textbf{0.82} & \textbf{0.82} & 0.53 & 0.56 & 0.47 & 0.64 & 0.55 & \textbf{0.60} & 0.27 & 0.37 & 0.38 & 0.50\\
			\hline
			65 & 0.79 & \textbf{0.81} & 0.54 & 0.57 & 0.49 & 0.54 & \textbf{0.68} & 0.65 & 0.19 & 0.41 & 0.43 & 0.31\\
			\hline
			70 & \textbf{0.85} & \textbf{0.85} & 0.43 & 0.57 & 0.51 & 0.66 & 0.67 & \textbf{0.75} & 0.19 & 0.50 & 0.49 & 0.50\\
			\hline
			75 & 0.83 & \textbf{0.85} & 0.41 & 0.63 & 0.53 & 0.77 & 0.48 & \textbf{0.56} & 0.07 & 0.56 & 0.45 & 0.15\\
			\hline
			80 & 0.81 & \textbf{0.83} & 0.29 & 0.61 & 0.45 & 0.81 & \textbf{0.12} & 0.11 & 0.00 & 0.00 & 0.00 & 0.00\\
			\hline
			85 & 0.78 & \textbf{0.85} & 0.14 & 0.74 & 0.36 & 0.81 & 0.28 & \textbf{0.34} & 0.19 & 0.00 & 0.00 & 0.04\\
			\hline
			90 & \textbf{0.30} & 0.24 & 0.00 & 0.00 & 0.00 & 0.02 & 0.16 & 0.19 & 0.06 & 0.00 & 0.00 & \textbf{0.48}\\
			\hline
			95 & 0.44 & 0.52 & 0.11 & 0.00 & 0.00 & \textbf{0.78} & 0.27 & 0.30 & 0.00 & 0.00 & 0.00 & \textbf{0.95} \\
			\hline
			98 & 0.49 & 0.57 & 0.00 & 0.00 & 0.00 & \textbf{0.76} & \textbf{0.42} & \textbf{0.42} & 0.01 & 0.00 & 0.00 & 0.21\\
			\hline
			MEAN & 0.69 & \textbf{0.71} & 0.30 & 0.42 & 0.33 & 0.58 & 0.42 & \textbf{0.45} & 0.13 & 0.22 & 0.21 & 0.28 \\
			\hline
		\end{tabular}
	}
	\caption{Accuracies obtained by the proposed approach compared to Bianchi et al. (\cite{bianchi2012image}), Galvan et al. (\cite{galvan2014first}), Dalmia et al. (\cite{dalmia2016first}) and Niu et al. (\cite{8945385}) with different combinations of $QF_1$/$QF_2$, considering standard quantization tables.}
	\label{tab:standard_tables}
\end{table}

\begin{table}[t]
	\centering
	\resizebox{\linewidth}{!}{
	\renewcommand{\arraystretch}{1.3}
		\begin{tabular}{c||c||c||c||c||c||c||c||c||c||c}
			\hline
			 \textbf{$PS$} & \multicolumn{5}{c ||}{\textbf{$QF_2=90$}} & \multicolumn{5}{c}{\textbf{$QF_2=80$}} \\
			\hline \hline
			 & \textit{Our} & \textit{Our Reg.} & \textit{\cite{bianchi2012image}} & \textit{\cite{galvan2014first}} & \textit{\cite{8945385}}   	& \textit{Our} & \textit{Our Reg.} & \textit{\cite{bianchi2012image}} & \textit{\cite{galvan2014first}}  & \textit{\cite{8945385}} \\
			\hline \hline
			5 & \textbf{0.80} & 0.78 & 0.56 & 0.58 & 0.05 & 0.65 & \textbf{0.68} & 0.26 & 0.46 & 0.07\\
			\hline
			6 & \textbf{0.82} & \textbf{0.82} & 0.46 & 0.60 & 0.07 & 0.42 & \textbf{0.54} & 0.05 & 0.41 & 0.02 \\
			\hline
			7 & \textbf{0.83} & \textbf{0.83} & 0.41 & 0.58 & 0.07 & 0.62 & \textbf{0.68} & 0.15 & 0.48 & 0.08 \\
			\hline
			8 & \textbf{0.81} & \textbf{0.81} & 0.25 & 0.65 & 0.10 & 0.19 & \textbf{0.22} & 0.03 & 0.03 & 0.01\\
			\hline
			9 & 0.55 & \textbf{0.61} & 0.02 & 0.47 & 0.02 & 0.26 & \textbf{0.28} & 0.19 & 0.00 & 0.07 \\
			\hline
			10 & 0.42 & \textbf{0.50} & 0.19 & 0.00 & 0.25 & 0.15 & \textbf{0.20} & 0.00 & 0.00 & 0.40 \\
			\hline
			11 & 0.45 & 0.52 & 0.04 & 0.00 & \textbf{0.69} & 0.37 & \textbf{0.38} & 0.01 & 0.00 & 0.24\\
			\hline
			12 & 0.49 & 0.57 & 0.04 & 0.00 & \textbf{0.75} & \textbf{0.42} & \textbf{0.42} & 0.01 & 0.00 & 0.21 \\
			\hline
			MEAN & 0.64 & \textbf{0.68} & 0.25 & 0.36 & 0.25 & 0.39 & \textbf{0.42} & 0.09 & 0.18 & 0.14\\
			\hline
		\end{tabular}
	}
	\caption{Accuracies obtained by the proposed approach compared to Bianchi et al. (\cite{bianchi2012image}), Galvan et al. (\cite{galvan2014first}) and Niu et al. (\cite{8945385}) employing custom tables for first compression. The column \textbf{$PS$} refers to custom tables used by Photoshop.}
	\label{tab:custom_tables}
\end{table}

\begin{figure}[t]
\centering
\captionsetup[subfloat]{labelformat=simple , labelsep=period}

     \subfloat[$QF_1:\{55,60,65,70,75,80,85,90,95,98\}$, $QF_2:90$ \label{subfig:res_std_90}]{%
       \includegraphics[width=\linewidth]{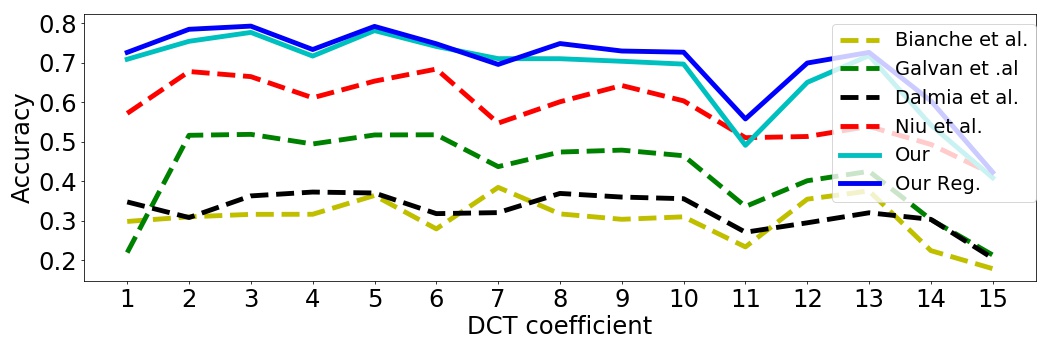}
     }
     \hfill
     \subfloat[$QF_1:\{55,60,65,70,75,80,85,90,95,98\}$, $QF_2:80$\label{subfig:res_std_80}]{%
      \includegraphics[width=\linewidth]{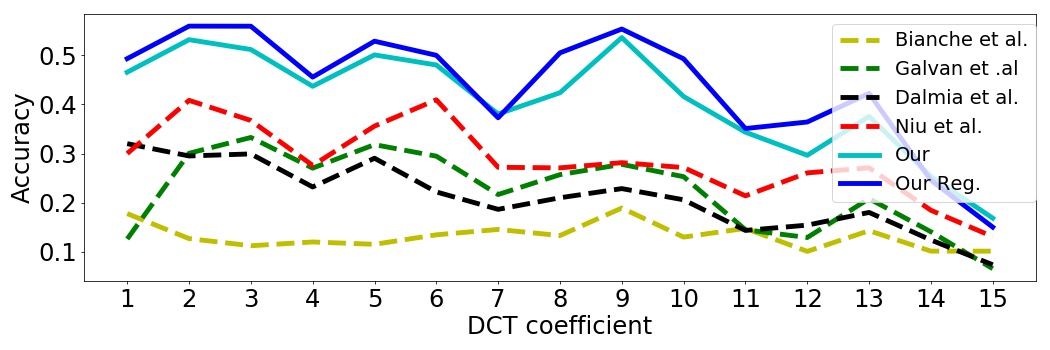}
     }
     \hfill
     \subfloat[$Q_{1}:\{5,6,7,8,9,10,11,12\}$, $QF_2:90$ \label{subfig:res_ps_90}]{%
       \includegraphics[width=\linewidth]{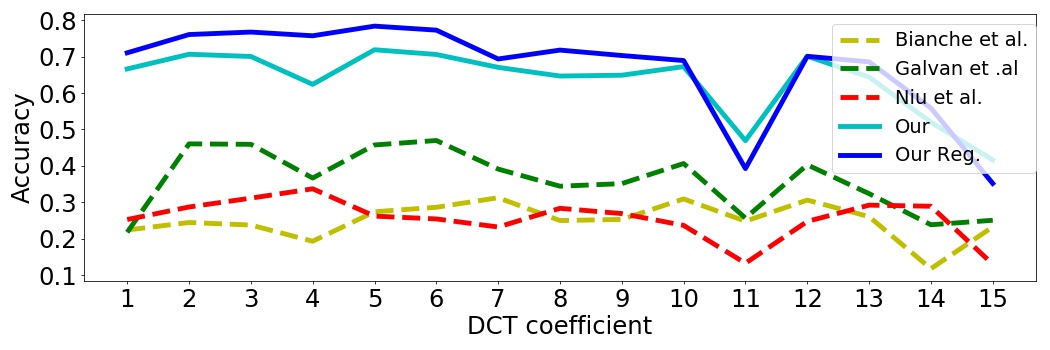}
     }
     \hfill
     \subfloat[$Q_{1}:\{5,6,7,8,9,10,11,12\}$, $QF_2:80$ \label{subfig:res_ps_80}]{%
      \includegraphics[width=\linewidth]{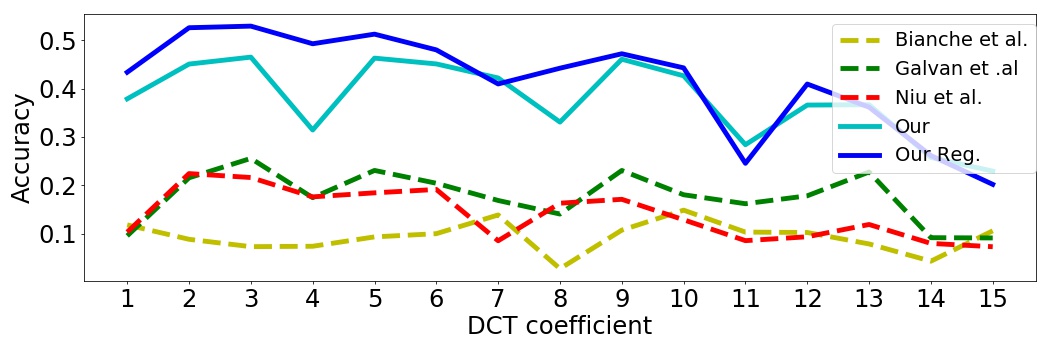}
     
     }

     \caption{Accuracies of the same methods described in Table \ref{tab:standard_tables} and \ref{tab:custom_tables} at varying of the quantization factors $q1_i$ to be predicted. The values are averaged over all the $QF_1/Q_1$.}
     \label{fig:accuracy_coefficients}
   \end{figure}

\label{section:comaprison_test}
The first set of tests has been performed to compare the proposed solution with state-of-the-art approaches based on statistical analysis (Bianchi et al. \cite{bianchi2012image}, Galvan et al. \cite{galvan2014first}, Dalmia et al. \cite{dalmia2016first}) and Machine Learning (Niu et al. \cite{8945385}), employing the implementations provided by the authors. 

For the sake of comparison, we created $4$ datasets: starting from RAISE \cite{dang2015raise}, we cropped a random $64 \times 64$ patch from every image and compressed it two times as follows:
\begin{enumerate}
\small
    \item $QF_1:\{55,60,65,70,75,80,85,90,95,98\}$, $QF_2:90$
    \item $QF_1:\{55,60,65,70,75,80,85,90,95,98\}$, $QF_2:80$
    \item $Q_{1}:\{5,6,7,8,9,10,11,12\}$, $QF_2:90$ 
    \item $Q_{1}:\{5,6,7,8,9,10,11,12\}$, $QF_2:80$ 
\end{enumerate}
where $Q_{1}:\{5,6,7,8,9,10,11,12\}$ related to 3) and 4) are referred to the quantization matrices of Photoshop (CC version 20.0.4). Every method was tested with all the aforementioned datasets with the exception of Dalmia et al. \cite{dalmia2016first} which, in their implementation, make assumptions about standard tables in first compression and then was excluded in the test with Photoshop's custom tables. Results, reported in Table \ref{tab:standard_tables} and \ref{tab:custom_tables} and in Fig. \ref{fig:accuracy_coefficients}, clearly highlight how our method outperforms the state-of-the-art in almost all scenarios, with and without the regularization step (values close to $0$ are due to the assumptions of some methods, e.g. $QF_1<QF_2$). It is worth noting that, although the scenario employing standard matrices is the one considered by Niu et al. \cite{8945385} to train their CNN, the proposed solution outperforms it with an average of \textbf{0.71} vs. \textbf{0.58} for $QF_2=90$ and \textbf{0.45} vs. \textbf{0.28}  for $QF_2=80$. Differently than other Machine Learning based approaches, the proposed method, working properly also in the scenario involving Photoshop's custom tables, demonstrates to be not dependent to a specific class of quantization matrices. 

\subsection{Discussion on unpredictable factors}
   
\begin{figure}[t]
\centering
     \hfill
     
      \includegraphics[width=\linewidth]{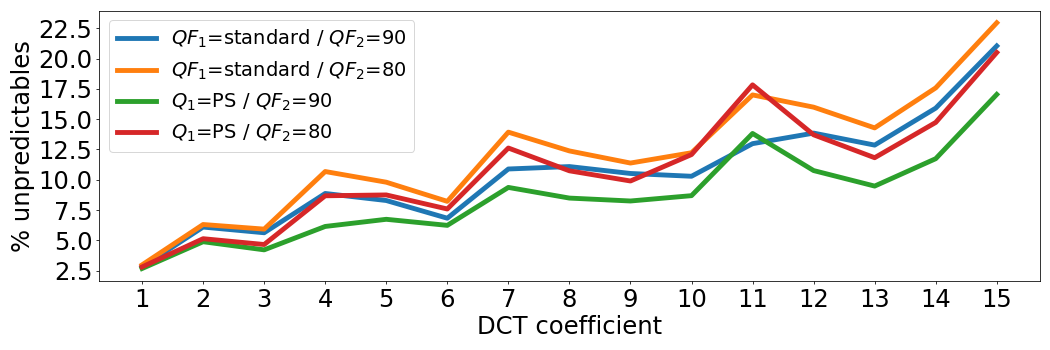}

     \caption{Percentages of unpredictables images for the $4$ described datasets at varying of DCT coefficients.}
     \label{fig:percentage_unpredictables}
   \end{figure}

A patch extracted from an high resolution image could be too homogeneous, hence all the values of the DCT coefficients could be in the same bin of the corresponding histograms, due to the lack of variation. A patch $64 \times 64$ extracted from a RAISE image represents about the $0,033\%$ of the information contained in the original one and then the possibility to have this homogeneous scenario 
is not negligible. We selected this scenario (i.e., small patches from an high resolution dataset) to test the proposed solution in a challenging condition, but we had to exclude the aforementioned histograms because they do not allow to predict the quantization factors (the available information is not enough to select a specific $q1$). In particular, given $i$, we exclude the distributions $h_i$ where all the values are in the same bin of the histogram (e.g., all the $i$-th DCT coefficients extracted form the $8\times8$ blocks are equal). The percentage of $q1_i$ excluded for each dataset is shown in Figure \ref{fig:percentage_unpredictables} at varying of DCT coefficients.

\subsection{Generalizing Property}

\begin{table*}[t]
	\centering
	\resizebox{\linewidth}{!}{
	\renewcommand{\arraystretch}{1.3}
		\begin{tabular}{c||c||c||c||c||c||c||c||c||c||c||c||c}
		\hline

		    \hline
		    \hline
		    
		    \textbf{Method} &\textbf{Dataset} &\textbf{Cropped Patch} &\textbf{Low/Low} & \textbf{Low/Mid} & \textbf{Low/High} &
		    \textbf{Mid/Low} & \textbf{Mid/Mid} & \textbf{Mid/High} &
		    \textbf{High/Low} & \textbf{High/Mid} & \textbf{High/High} & \textbf{Mean}\\
		    
		    \hline
			Our & \multirow{2}{*}{RAISE \cite{dang2015raise}} & \multirow{2}{*}{$64\times64$} & 0.25 &	0.47  & 0.79 & 0.17 & 0.32 & 0.82 & \textbf{0.27} & 0.31 & 0.70 & 0.46\\
			\cline{1-1}\cline{4-13}
			Our Reg. &  &  & \textbf{0.30} & \textbf{0.53} & \textbf{0.81} & \textbf{0.22} & \textbf{0.37} & \textbf{0.84} & 0.25 & \textbf{0.33} & \textbf{0.75} &  \textbf{0.49}\\
			\hline
			\hline
			Our & \multirow{2}{*}{UCID \cite{schaefer2003ucid}} & \multirow{2}{*}{$64\times64$} & 0.33 & 0.63 & 0.93 & 0.20 & 0.39  & 0.90 & \textbf{0.15} & 0.21 & 0.66 & 0.49\\
			\cline{1-1}\cline{4-13}
			Our Reg. &  &  & \textbf{0.36} & \textbf{0.65} & \textbf{0.96} & \textbf{0.23} & \textbf{0.42} & \textbf{0.91} & 0.13 & \textbf{0.23} & \textbf{0.73}  & \textbf{0.51}\\
			\hline
			\hline
			Our & \multirow{2}{*}{UCID \cite{schaefer2003ucid}} & \multirow{2}{*}{$128\times128$} & 0.36 & 0.69 & \textbf{0.95} & 0.21 & 0.42  & \textbf{0.92} & 0.16 & 0.24 & 0.71  & 0.52 \\
			\cline{1-1}\cline{4-13}
			Our Reg. &  &  & \textbf{0.37} & \textbf{0.71} & \textbf{0.95} & \textbf{0.24} & \textbf{0.43} & \textbf{0.92} & \textbf{0.24} & \textbf{0.33} & \textbf{0.74} &  \textbf{0.55}\\
			\hline
			\hline


		\end{tabular}
	}
	\caption{Accuracies obtained by the proposed approach for generalizing property demonstration (Our Reg. for the regularized version).}
	\label{tab:custom_tables_full_raise}
\end{table*}

Park et al. \cite{Park_2018_ECCV} proposed a collection of $1170$ different quantization tables employed on real scenarios ($1070$ custom). In this Section, a new set of experiments will be presented in order to demonstrate the generalizing property of the proposed approach employing the same reference dataset built in Section~\ref{sec:dataset}. 

From the collection of Park et al., $873$ quantization tables show in the first $15$ quantization factors, a value of $q1_i \leq q1_{max}$ and then they can be employed for testing. They were sorted by the average of the first $15$ quantization factors and then equally divided into three sets of $291$ elements (\textit{low}, \textit{mid}, \textit{high}). These sets of tables are employed to create $9$ combinations of double compressions (see latter $9$ columns of Table \ref{tab:custom_tables_full_raise}). 
Three new datasets for generalizing tests are then built: patches of size $w \times w$ from RAW images were extracted and then compressed two times with all the $9$ aforementioned combinations. For each combination of double compression, the quantization tables ($Q_1$ and $Q_2$) are selected randomly from the $291$ available in the corresponding set.

The results obtained on these three datasets are reported in Table \ref{tab:custom_tables_full_raise}. They clearly show that our method maintains good accuracies on all the tested challenging scenarios and demonstrates to achieve same results even when different datasets are employed for tests.

\section{Conclusion}
\label{sec:conclusion}
In this paper, a new technique able to estimate the first quantization factors for JPEG double compressed images was presented, employing a mixed statistical and Machine Learning approach.
One of the main contributions was the way we employed the big amount of data to avoid overfitting: constant matrices $M_i$ permitted to uncouple $\{q1,q2\}$ and the use of $\mu$ and $\beta$ made computational times acceptable. The presented solution was demonstrated to work for both custom and standard tables thus being generalizable enough to be employed in real-case scenarios. Experimental tests showed the goodness of the technique overcoming state-of-the-art results. Finally, the use of $1$-nn to learn the distribution underlines rooms for improvement of the proposed method.




%

%
%
%
%
%


\balance
{\small
\bibliographystyle{IEEEtran}
\bibliography{main}
}
\end{document}